# CryoDyna: Multiscale end-to-end modeling of cryo-EM macromolecule dynamics with physics-aware neural network

Chengwei Zhang[1,2,3#], Shimian Li[4#], Yihao Niu[4], Zhen Zhu[3,5], Sihao Yuan[4], Sirui Liu[3*], Yi Qin Gao[1,3,4*]

1. Biomedical Pioneering Innovation Center (BIOPIC), Peking University, Beijing 100871, China

2. School of Life Sciences, Peking University, Beijing 100871, China

3. Changping Laboratory, Beijing 102200, China

4. New Cornerstone Science Laboratory, Beijing National Laboratory for Molecular Sciences, College of Chemistry and Molecular Engineering, Peking University, Beijing 100871, China

5. Academy for Advanced Interdisciplinary Studies, Peking University, Beijing 100871, China

\# These authors contributed equally.

\* To whom correspondence should be addressed.

# Abstract

Single-particle cryo-EM has transformed structural biology but still faces challenges in resolving conformational heterogeneity at atomic resolution. Existing cryo-EM heterogeneity analysis methods either lack atomic details or tend to subject to overfitting due to image noise and limited information in single views. To obtain atomic detailed multiple conformations and make full use of particle images of different orientations, we present here CryoDyna, a deep learning framework to infer macromolecular dynamics directly from 2D projections by integrating cross-view attention and multi-scale deformation modeling. Combining coarse-grained MARTINI representation with atomic backmapping, CryoDyna achieves near-atomic interpretation of protein conformational landscapes. Validated on multiple simulated and experimental datasets, CryoDyna demonstrates improved modeling accuracy and robustly recovers multi-scale complex structure changes hidden in the cryo-EM particle stacks. As examples, we generated protein-RNA coordinated motions, resolved dynamics in the "unseen" region of RAG signal end complex, mapped translocating ribosome states in a one-shot manner, and revealed step-wise closure of a membrane-anchored protein multimer. This work bridges the gap between cryo-EM heterogeneity analysis and atomic-scale structural dynamics, offering a promising tool for exploration of complex biological mechanisms.

# Introduction

Single particle cryogenic electron microscopy (cryo-EM) has gained popularity for its structure resolving ability of biological macromolecules at near-atomic resolution in a time-efficient manner[1,2]. Averaging particle images of macromolecular complexes captured from different orientations in vitreous ice allows the reconstruction of a homogeneous model[3]. A potential limitation, however, lies in the inherent conformational heterogeneity of proteins and other molecules, which may result in locally diminished resolution and the loss of functionally relevant dynamic information[4]. Reconstructing the structural landscape of biological macromolecules from noisy 2D particle images represents an urgent challenge in the field.

Conventional conformational heterogeneity[5–8] analysis methods typically predict discrete states via iterative multi-volume optimization. These approaches are sensitive to initial volume quality and also require pre-defining cluster numbers, posing significant limitations on their applicability. Recent methods have attempted to leverage statistical analysis like Principal component analysis (PCA)-based methods[9–11] or deep learning techniques to decipher continuous conformational heterogeneity from cryo-EM datasets. For example, CryoDRGN[12] and CryoDRGN2[13] utilize neural networks to learn a non-linear latent space with a variational autoencoder architecture in Fourier domain to map conformational variations. While in voxel-based representation, E2gmm[14] approximates density maps using a set of learnable Gaussian distributions, and 3DFlex[15] jointly optimizes a consensus map and deformation fields in a mesh representation. These methods are effective in capturing large-scale conformational changes but fall short in providing atomic-level explanations for the predicted structural variants, primarily as a consequence of the image-to-volume training scheme.

Integrating structure modeling into the neural networks can provide direct interpretation and constraint the deformation to a sensible space. Some methods use normal mode analysis (NMA)[16] or Zernike polynomials[17] to model heterogeneity with a combination of bases, but they are limited to resolve either small-scale continuous motion or global deformation due to the limited expressive power of the base function. To take full advantage the power of deep learning, CryoSTAR[18] directly predicts per-residue motion and introduces structural regularization derived from the reference structure to keep local rigidity. Nevertheless, CryoSTAR's residue-independent deformation prediction makes it susceptible to noise overfitting. CryoSPHERE[19] addresses rigid-body heterogeneity via joint optimization of structural segmentation and image-specific transformations, yet struggles with complex datasets. A fundamental limitation of both methods lies in their 'one-image-one-conformation' training framework, and the lack of multi-view supervision for the same potential conformation compromises their prediction reliability.

Here, we present CryoDyna, a structure-aware deep-learning tool for modeling macromolecule dynamics by inferring deformation fields directly from noisy 2D projections. CryoDyna introduces a novel cross-view attention module to fuse information from images with different orientations, as well as a multi-scale deformation prediction module to capture the coordinated motions of residues. To provide more detailed atomic interpretation, we expand the modeling entity to coarse-grained (CG) beads defined by MARTINI[20–23], enabling the exploration of conformational heterogeneity at near-atomic scale, which can be further backmapped to atomic level. We validate the performance of CryoDyna on two simulated cryo-EM datasets[18,24] and four real experimental datasets[25–28]. CryoDyna robustly resolves conformational states that substantially deviate from reference structures on both simulated datasets. The CG bead-level reconstruction framework further ensures accurate and physically sound all-atom structure recovery. CryoDyna successfully recapitulates the structural dynamics identified by existing methods in the yeast spliceosomal U4/U6.U5 tri-snRNP[25] (EMPIAR-10073) complex, and infers physically reasonable coordinated motions between proteins and RNAs. For the RAG signal end complex[26] (EMPIAR-10049), CryoDyna reconstructs the structural dynamics of unresolved regions coupled with AlphaFold3[29] (AF3) modeling. CryoDyna can also automatically identify major functional states during ribosomal translocation[27] (EMPIAR-10792). Finally, we show the performance of CryoDyna on modeling the stepwise closure of membrane-anchored protein complexes, AP-1:Arf1:tetherin-Nef trimer[28] (EMPIAR-10177). These results are cross-validated by training a volume decoder on the same learnt latent space. Our experiments show that CryoDyna can help effectively and accurately explore hidden structural dynamics in cryo-EM particle stacks yet to be characterized at residue to atomic scales in a physics-informed manner.

# Results

**The physics-aware multiscale CryoDyna method**

The neural network in CryoDyna consists of a multi-view image encoder and a hierarchical deformation decoder (Fig. 1a). Different from previous methods which predict conformations only using corresponding images, CryoDyna utilize a cross-view attention mechanism to explicitly extract information across different images in the same batch. In this way, particle images in the same orientation will be aggregated to remove random noise and particle images in different orientations can help to improve geometric consistency. Each particle image $I \in \mathbb{R}^{D \times D}$ from the dataset is converted into a compact representation $Z \in \mathbb{R}^C$ through this image encoder.

To simulate the hierarchical movement of protein complexes, a multi-scale deformation decoder merge and broadcast information bidirectionally between residue-level and secondary structural elements (SSE)-level representations (Fig. S1). This strategy enables the network to learn large-scale motions in the early stage and refine local structure in the latter stage. The updated residue-level representation is used to predict relative deformation $\Delta S$ against an initial reference structure $S \in \mathbb{R}^{N \times 3}$. A simulated density $V$ derived from predicted deformation then passes through the image forward model and generates corresponding 2D projections. The correlation between input image and predicted 2D projection supervise learnt motion (Fig. 1a). CryoDyna also adopts structural regularization terms to keep structural validity (Methods).

To resolve atomic-detailed structures beyond residue-level resolution, we further integrate the coarse-grained MARTINI force field (version 2.2) into CryoDyna, hereafter referred to as CryoDyna-CG. The MARTINI-based CG parameterization[20–23] is utilized in both the model architecture and the structure regularization, balancing between structural detail and computational efficiency (Fig. 1b; Methods). In this framework, MARTINI-mapped CG structures act as structural priors for predicting relative deformations, with each CG bead representing about 4 heavy atoms on average. This strategy raises the theoretical upper bound of approximation accuracy for cryo-EM density and projection to near-atomic level. The MARTINI force field parameters, which describe the physical interactions between beads, are then incorporated as structural regularization to guide physically plausible conformational variability during heterogeneous reconstruction. We employ bonded interactions including bond, angle, and dihedral terms to constrain the relative spatial arrangement between beads, as well as nonbonded interactions including Morse potentials and an elastic network (EN) term to respectively prevent bead clashes and maintain local rigidity. A combination of CG2AT2[30] and Backward[31] backmapping methods is employed to convert coarse-grained models into all-atom models, enabling more detailed structural insights at the atomic scale (Fig. 1c).

Upon completion of model training, the learned latent space enables systematic exploration of structural heterogeneity within the dataset, facilitating subsequent key applications: (1) generation of continuous conformational transition pathways, (2) fast discovery of structural states, and (3) quantitative analysis of state populations. The back-mapped all-atom structures provide highly detailed insights to directly bridge cryo-EM particle image with molecular dynamics simulations, thus establishing a comprehensive framework for mechanistic studies of protein complexes.

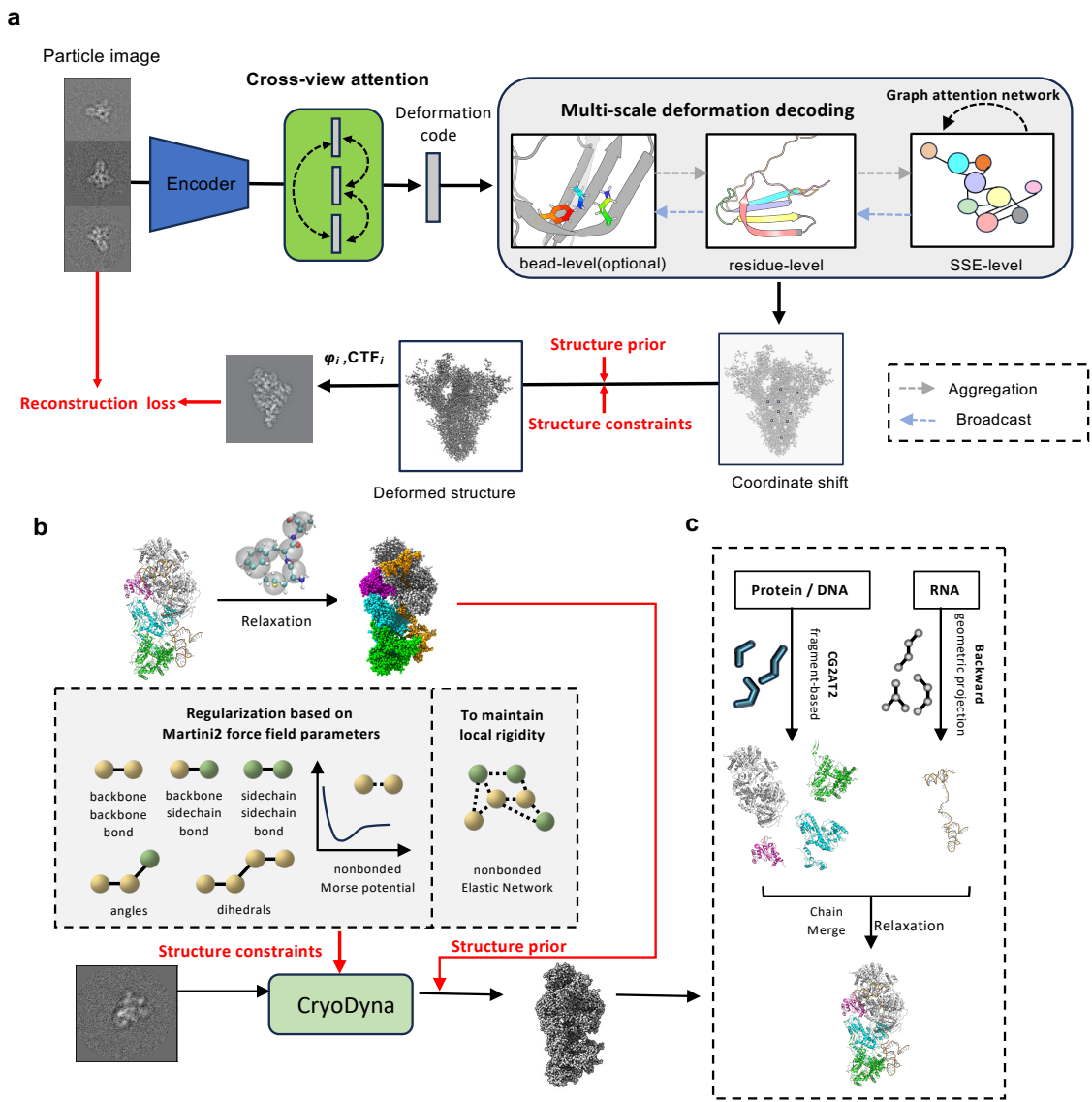

**Fig. 1 | CryoDyna resolves multi-scale conformational heterogeneity with atomic details. a,** The schematic figure of CyroDyna. CryoDyna method adopts a cross-view attention mechanism for images in the same batch and predicts the coordinate shift relative to the reference structure through a multi-scale deformation decoder. The model training is supervised by the correlation between the input images and the predicted 2D projections from the deformed structure with given pose and CTF. Additionally, a structural regularization term containing continuity loss, clash loss and elastic network loss is incorporated into the loss function to keep structural validity. **b,** Integration of MARTINI force field in CryoDyna-CG. For bead-level prediction, the MARTINI-mapped coarse-grained initial structure after energy minimization is used as structure prior. Bonded terms (bond, angle, dihedral) and nonbonded Morse term derived from MARTINI parameters are used as structural constraint losses, and Elastic Network term derived from the coarse-grained initial structure are used as structural prior losses. **c,** Backmapping procedure for all-atom model construction in CryoDyna-CG. Protein and DNA chains are reconstructed into all-atom models using the fragment-based CG2AT2 method, whereas RNA chains are converted with the Backward approach. The reconstructed chains are then merged and subjected to position-restrained equilibration, yielding the final all-atom structure.

# CryoDyna robustly recovers distant conformational states on simulated datasets

To evaluate CryoDyna's capability in modeling conformational heterogeneity, we benchmarked its performance using two simulated datasets with known ground-truth structures. The first dataset, derived from Ref.[18], simulates the one-dimensional transition of Escherichia coli adenylate kinase between its closed (PDB:1AKE) and open (PDB:4AKE) states through structure interpolation. The dataset comprises 50,000 particle images, with 1,000 projections per interpolated structure. Using the closed state as the initial conformation, we compared the reconstruction performance of CryoDyna and CryoSTAR, another representative structure integrated heterogeneous reconstruction method. CryoDyna successfully reconstructed the complete transition pathway from the closed to open state (Fig. 2a) and displayed a faster convergence speed with a mean Cα-RMSD of 1.148 Å at 10th epoch, while CryoSTAR achieved a mean Cα-RMSD 1.497 Å at the same epoch (Fig. 2b). After 20 epochs, CryoDyna maintained superior reconstruction quality (mean Cα-RMSD: 1.110 Å vs. CryoSTAR's 1.475 Å). When structures were ranked by their deviation from the initial state, CryoDyna exhibited enhanced accuracy in resolving conformations distinct from the closed state, outperforming CryoSTAR across the spectrum of conformational heterogeneity (Fig. 2c).

The second simulated dataset was sampled from a molecular dynamics (MD) trajectory of the SARS-CoV-2 spike protein (3438 residues, Ref.[24]). This dataset covers the transition of the receptor binding domain (RBD) from the three-down to the one-up state, simulating both opening and twisting motions[32]. Using the seed structure in one-up state as the initial reference structure, we benchmarked CryoDyna on this dataset. As shown by the representative conformation sampled from the first principle component (PC1) of the latent space (Fig 2d), CryoDyna successfully generated physically realistic intermediates along the transition path toward the three-down state, demonstrating its ability to model large-scale conformational changes. CryoDyna again achieved a lower mean Cα-RMSD (3.62 Å) than CryoSTAR (4.26 Å) across 1,000 sampled images after 80 training epochs (Fig. 2e). In addition, CryoDyna consistently outperformed CryoSTAR in resolving various states, particularly for structures far from the initial one-up state. For example, CryoDyna correctly predicted a representative conformation near 400th frame in the simulation dataset that takes three-down conformation, while CryoSTAR erroneously generated a hallucinated one-up conformation. This observation highlights the importance of CryoDyna's cross-view attention mechanism, which mitigates single-image conformation hallucination.

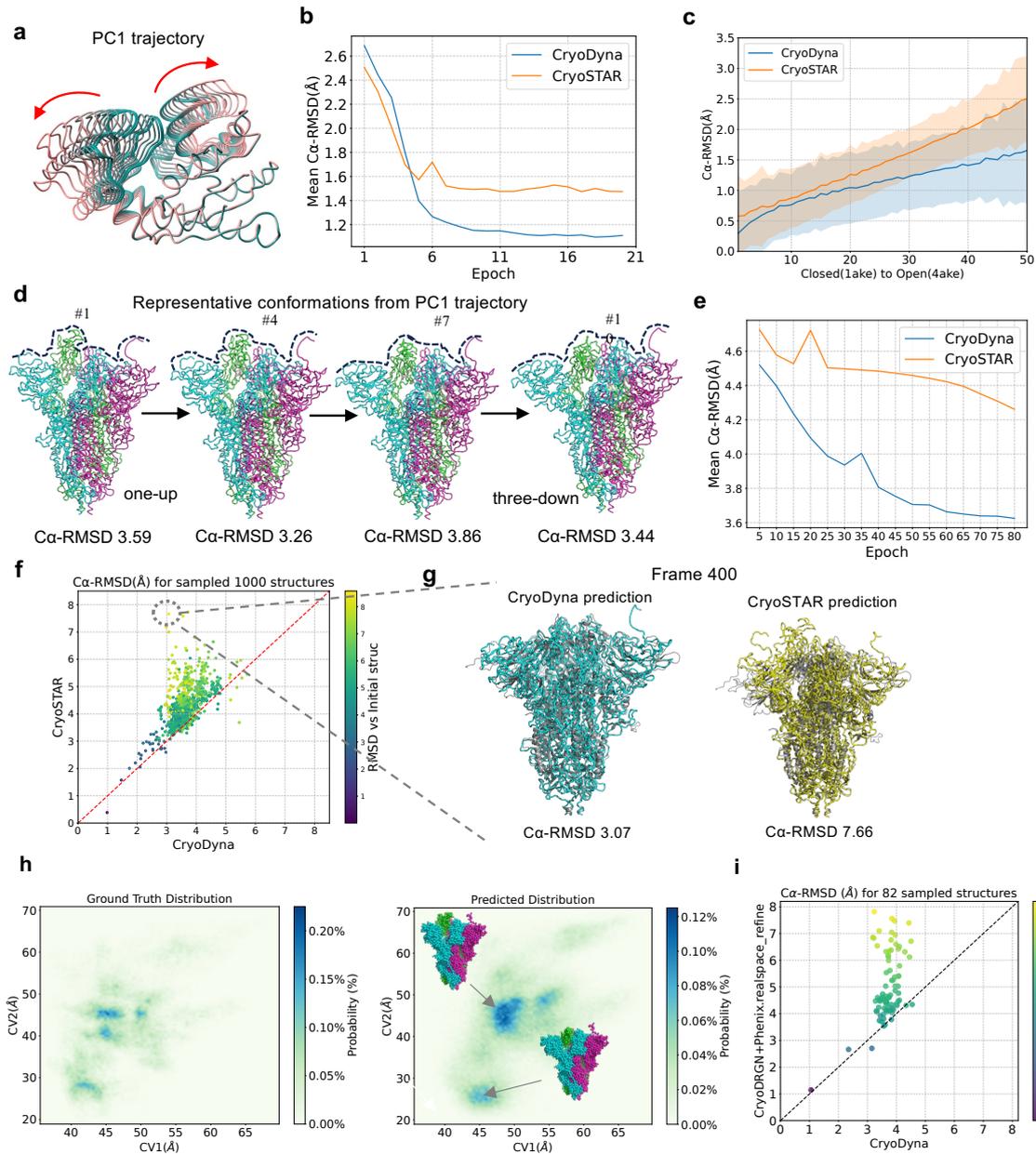

**Fig. 2 | Validating CryoDyna on two simulated datasets. a,** Ten representative structures sampled along PC1 of the latent space trained on adenylate kinase dataset. **b,** Mean Cα-RMSD of all 50,000 structures in different training epochs for CryoDyna (blue) and CryoSTAR (orange). **c,** Cα-RMSD of each conformation from closed state (left) to open state (right) for both methods. The middle curve refers to the mean values and the band regions represents one sigma deviation. **d,** Four representative structures sampled along PC1 of the latent space trained on the spike dataset. **e,** Mean Cα-RMSD of sampled 1,000 structures in different training epochs from CryoDyna (blue) and CryoSTAR (orange). **f,** Scatter plot of the Cα-RMSD for sampled 1,000 structures constructed by CryoDyna and CryoSTAR. The colorbar denotes the Cα-RMSD between target conformations and initial conformation. **g,** Predicted structures for the 400th frame of MD trajectory labelled in **f** from CryoDyna (paleblue) and CryoSTAR (yellow). **h,** Bead-level conformational distribution of ground truth (left) and that generated by CryoDyna (right). CV1 and CV2 correspond to the distances between the Backbone beads of A499–B116 and A499–B406, respectively. **i,** Comparison of Cα-RMSD against the ground truth structures of 82 sampled structures reconstructed by CryoDyna and CryoDRGN plus Phenix refinement.

## CryoDyna-CG enables all-atom structural interpretation for conformational heterogeneity

The integration of MARTINI-based coarse-grained structural representation and regularization into CryoDyna enables structural reconstruction at a resolution higher than the residue level. Combined with backmapping methods, this approach can further allow the recovery of all-atom structures. Since the backmapping procedure involves conformational energy minimization and NVT simulations, successful backmapping itself indicates that the conformation does not contain significant unphysical conformations or interactions. We validated its performance on the two simulated datasets.

For the adenylate kinase, the 214-residue system was mapped to 450 CG beads after MARTINI parameterization. At the 10th epoch, the average bead-RMSD between the generated structure and the reference structure (coarse-grained with MARTINI) was 2.00 Å, which further decreased to 1.49 Å by the 30th epoch (Fig. S2a). In the predicted structures, the distances between bonded beads were highly consistent with the equilibrium bond lengths specified by the Martini2 force field, with an average deviation of only -0.0640 Å (Fig. S2b). Further backmapping of 500 sampled CG structures yielded all-atom models with an average Cα-RMSD of 1.43 Å and an all-atom-RMSD of 1.85 Å (Fig. S2c).

For the SARS-CoV-2 spike protein, we applied a similar procedure to achieve high-resolution structural reconstruction. By the 40th epoch, the average bead-RMSD for sampled 1000 structures between the generated and reference structures was 4.36 Å, which further decreased to 4.22 Å by the 80th epoch (Fig. S3a). The mean deviation between bonded bead distances and their equilibrium lengths was only −0.0505 Å (Fig. S3b), indicating high structural fidelity. To evaluate whether CryoDyna-CG can enable reliable ensemble analysis, we analyzed the distance distributions of two residue pairs that characterize the motion of the RBD from the one-up to three-down state. The distributions of CryoDyna-CG conformations closely match the ground truth (Fig. 2h). All 100 sampled CG structures were successfully backmapped (Fig. S3c), yielding all-atom models with an average Cα-RMSD of 3.71 Å and an all-atom RMSD of 4.49 Å. The performance of CryoDyna-CG's end-to-end all-atom reconstruction is then compared with a two-stage alternative pipeline (Methods), which uses CryoDRGN to obtain density maps representing multiple conformations, and Phenix[33] to refine initial models in real space into the target conformations. Among the sampled 100 density maps reconstructed by CryoDRGN, only 82 can be successfully refined with Phenix and the refined models showed much lower quality compared to CryoDyna-CG (Ca-RMSD: 5.03 Å vs 3.70 Å;

Fig. 2i; Fig. S3d). This comparison highlights the advantage of the end-to-end structural reconstruction used in our model.

## CryoDyna captures reasonable coordinated motion of protein and RNA in the yeast spliceosomal U4/U6.U5 tri-snRNP

U4/U6.U5 tri-snRNP is an important component of fully assembled spliceosome and exhibits rich structural flexibility except for its core body[34]. We analyzed a dataset (EMPIAR-10073) of yeast spliceosomal U4/U6.U5 tri-snRNP (32 protein chains and 3 RNA chains with 9325 residues or nucleotides in total) from Nguyen et al.[25] using CryoDyna and recovered conformational heterogeneity across different parts and coordinated motion of snRNPs. We utilized the atomic structure (PDB:5GAN) resolved by consensus reconstruction[25] as input and trained CryoDyna on the full dataset with 138,899 particle images.

We sampled representative structures along the first and second principal components (PC1 and PC2) of the CryoDyna latent space (Fig. 3a). Notably, the sampled structures along PC1 component reveal an apparent motion between the head and foot domains, leading to open and close conformations (Fig. 3b). The sampled trajectory along PC2 captured the motion of head domain which flipped out-of-plane (Fig. 3c). Above predictions are in good agreement with the global and masked classification result by previous study[25]. In the PC2 trajectory, we additionally found a coupled motion between RNA and protein. The U4/U6 RNAs undergoes a slight rotational movement in coordination with the rotation of the arm domain (Fig. 3d, Supplementary Video 1).

In contrast, although CryoSTAR also discovered motions in the head, foot and arm domains (Fig. S4a), its predicted deformation and volume change of the arm domain contradicts with the fixed-size reconstruction result using standard masked refinement protocol[25] (Fig. S4b). Moreover, the rotation of the arm domain compresses the U4/U6 RNA at positions 70–88, resulting in distorted nucleic acid geometry (Fig. S4c).

To evaluate whether the predicted deformations by CryoDyna are reasonable, we trained a volume decoder using the learned latent variables to check the consistency of the predicted structures and density maps (Methods). The density maps generated by the volume decoder effectively cover the corresponding structural regions (e.g., the arm domain) and exhibit strong overall consistency (Fig. 3e). However, the arm domain shows a slightly lower consistency. Since the density maps produced by this method do not provide sufficient resolution to validate the dynamic behavior of U4/U6 RNAs, further experimental verification is desired for assessing the motion of RNA components.

Reconstruction by CryoDyna-CG further validates the above observations (Fig. S5). For example, a clear transition of the head and foot domains from an "open" to a "closed" state and cooperative motions between the arm domain and RNA were reproduced (Supplementary Video 2). Furthermore, we performed a backmapping focused on the arm regions of the conformations sampled along PC2. This analysis clearly showed that the arm domain rotates around a hinge formed by the Prp3, the LSM complex subunit LSM2, and the U6 snRNA (Fig. 3f). The heterogeneity captured by CryoDyna-CG after backmapping is not only consistent with the residue level observations, but also provides coordinated molecular motion details on this functionally important protein-RNA complex.

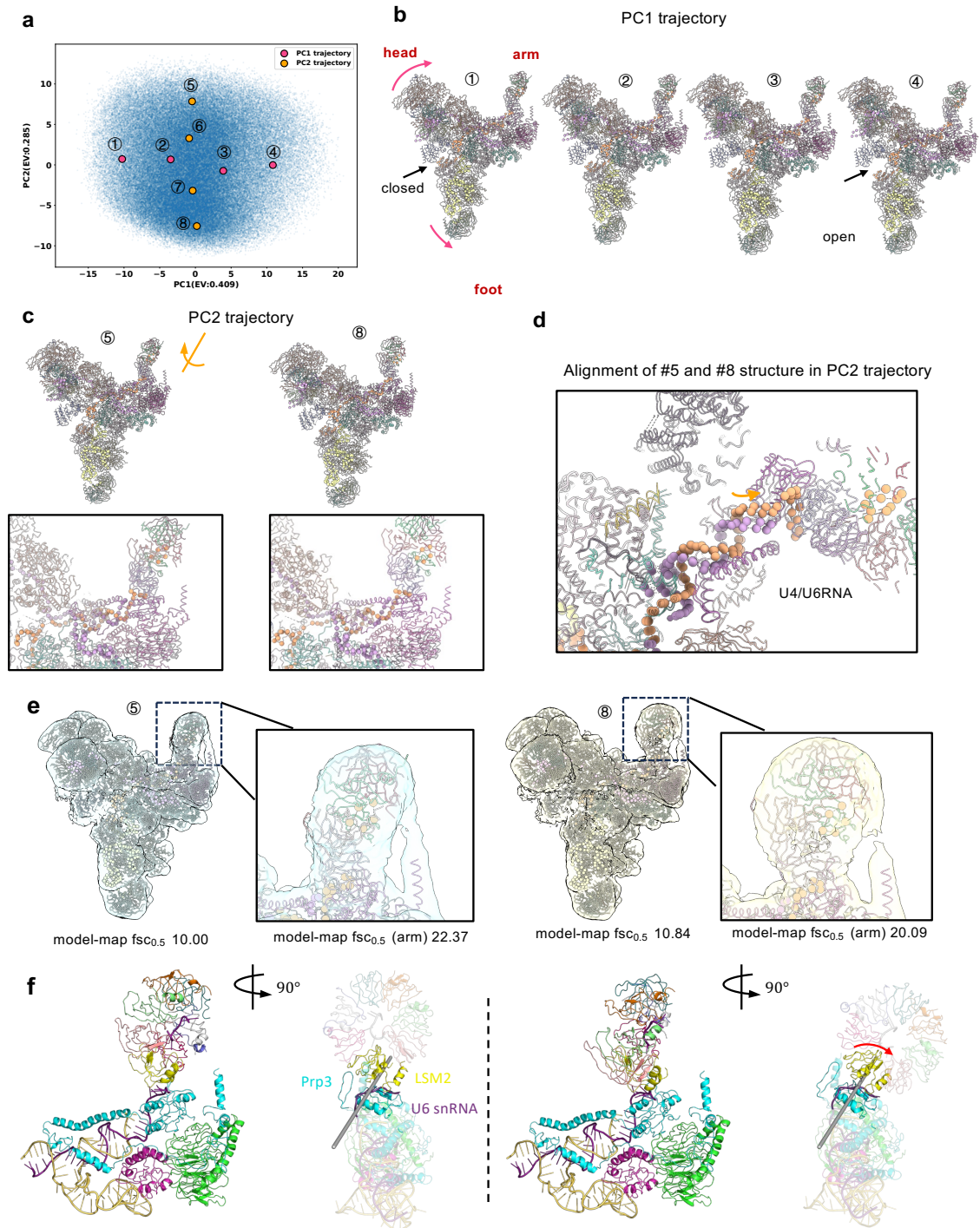

**Fig. 3| CryoDyna reveals reasonable coordinated protein-RNA motion in the yeast spliceosomal U4/U6.U5 tri-snRNP (EMPIAR-10073). a**, PCA visualization of the cryoDyna latent space. EV, explained variance. Four representative conformations sampled along PC1 are colored in red while four representative conformations sampled along PC2 are colored in yellow. **b,** Four representative conformations labeled as 1-4 in **a** sampled along PC1 of the latent space. **c**, Two representative conformations labeled as 5, 8 in **a** along PC2 of the latent space. The down two panels zoomed in to show the details of the arm domain. **d**, Alignment of 5th and 8th structures labeled in **a** shows minimal rotation of U4/U6 RNA. **e**, Superposition of the 5th (left) and 8th (right) structures with the corresponding predicted density maps by volume decoder. The right side of each panel zoomed in to show alignment details of atomic structures and density maps. Model-map $FSC_{0.5}$ measures the consistency between the atomic structure and the density map. **f,** Backmapped all-atom structures of the arm domain. The left and right panels correspond to conformations #3 and #5 from the PC2 trajectory shown in Fig. S5a, respectively. The gray stick represents the pivot.

## CryoDyna provides structural insight for unresolved regions in RAG signal-end complex

The RAG complex (RAG1/RAG2) precisely cleaves recombination signal sequences (RSS) flanking V, D, and J gene segments, thereby generating the combinatorial diversity essential for adaptive immunity[35]. Ru et al.[26] resolved RAG complex in two distinct stages (Fig 4a): a signal-end complex (SEC) lacking distal RSS ends and NBD density (EMD-6487), and a well-resolved non-symmetrized paired complex(PC) enabling atomic modeling of these DNA-protein interaction sites (EMD-6489). We evaluated the ability of CryoDyna to provide structural dynamics for the unresolved regions of the signal-end complex (SEC) within this dataset, specifically focusing on the distal 12/23-RSS termini, RAG1's nonamer binding domain (NBD), RING domain and zinc finger A domain (RING ZnA) that were missing in the original cryo-EM reconstruction[26].

We specifically noted that the RAG complexes were reconstituted by incubating the zRAG1-RAG2 complex, 12-RSS, 23-RSS, and HMGB1 in a 2:1:1:3 molar ratio, so three HMGB1 boxes were included when predicting the reference structure with AF3. The overall structure predicted by AF3 largely matches the asymmetric PC complex, except for some discrepancies in the bending angles of the 12-RSS and 23-RSS, as well as the absence of density for the Ring ZnA domain (Fig. 4a ,middle panel). Analysis of the electrostatic surface in the AF3-predicted structure suggests that it reasonably generates a polar DNA-binding path (Fig. 4a, right panel). We used this approximate structure to train CryoDyna for SEC dynamics exploration.

We sampled representative structures along the PC1 trajectory in latent space (Fig. 4b) and observed that the 23-RSS undergoes additional bending compared to the crystal structure while the 12-RSS straightens from its bent conformation (Fig. 4c，Fig. S6a, Supplementary Video 3). We validated these conformations by training a volume decoder with the learnt latent representation as input. Surprisingly, the reconstructed density map revealed multiple structural components that were partially or completely unresolved by previous reconstruction methods like CryoDRGN[12] (Fig. S6b), including the distal 12-RSS and 23-RSS elements, NBD dimer interface, HMG box domains, and the RING-ZnA domain of RAG1 (Fig. 4d). This suggests CryoDyna may have learned a conformational distribution that effectively distinguishes different structural states, thereby enhancing signals corresponding to respective conformations in regions that would otherwise lack density. The observed consistency between cryo-EM densities and predicted structures (Fig. 4d; Fig. S6a) indicates the reliability of CryoDyna.  In particular, 12-RSS and 23-RSS exhibit coordinated anti-phase rotations relative to the core region (Fig. 4e, left panel). These RSS components function as two arms of a molecular lever formed by RAG1's NBD dimer (scaffold) and RING-Zn domain (mechanical arm): when one rotates inward toward the core, it drives synchronous outward rotation of the other (Fig. 4e, right panel). This motion pattern was also observed in the 3D classification results of RAG paired complex[26].

Reconstruction by CryoDyna-CG exhibited motions similar to the above observations except that the 23-RSS bending was less pronounced (Fig. S7a; Supplementary Video 4). Backmapping of ten representative structures showed that most base-pair parameters (Shear, Stretch, Stagger, Buckle, Opening) matched the experimental structure (PDB: 3JBW), indicating that base-pair geometry preserved normal hydrogen-bonding and stacking interactions (Fig. S7b). Propeller angles do display anomalous positive values in local regions such as 16-19 nucleotides in 23 RSS, likely due to local DNA adaptation upon protein binding (Fig. S7c). The base-step parameters Twist and Rise are both in good agreement with the resolved structure (Fig. S7d). Interestingly, the all-atom modeling helped to reveal the potential driving force underlying the "lever motion" of the RAG1 dimer: when the lever swings upward, new inter-chain interactions form between the RING ZnA domain and the loop region of the N-terminal domain (NTD) (e.g., hydrogen bonds such as D336-T65 and R330-N67, Fig. 4e, right panel), thereby stabilizing this conformation. The all-atom resolution thus provides more detailed and reliable structural insights into the dynamic mechanisms of the complex.

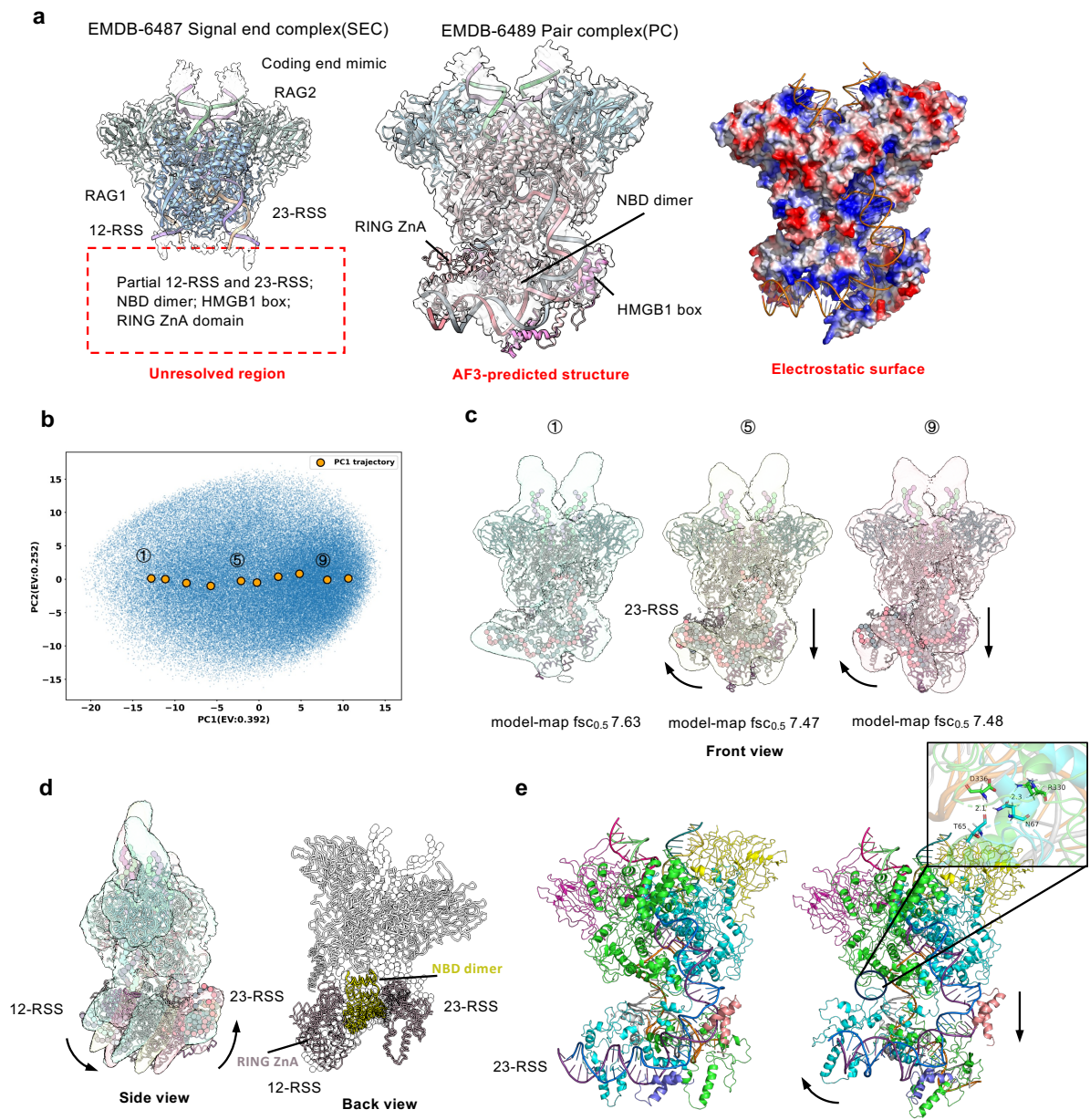

**Fig. 4| CryoDyna uncovers "unseen" dynamics of unresolved regions in RAG signal-end complex (EMPIAR-10049). a,** Left panel shows the superposition of experimental density map (EMDB-6487) and resolved structure (PDB:3JBX) of RAG signal-end complex. Middle panel shows the superposition of experimental density map (EMDB-6489) for RAG Pair complex and the atomic structure predicted by AlphaFold3 for the full system of RAG Signal-end complex. Right panel shows the electrostatic surface of the RAG dimer and HMG boxes from the AF3-predicted model with bound RSSs and coding end mimics in ribbons. **b,** PCA visualization of the cryoDyna latent space. EV, explained variance. Ten representative conformations sampled along PC1 are colored in yellow. **c,** Superposition of three representative conformations labeled as 1,5,9 (left to right) in **b** with their corresponding predicted density maps by volume decoder in front view. Model-map $FSC_{0.5}$ measures the consistency of the atomic structure and the corresponding density map. All density maps are shown using the same isosurface level. **d,** Superposition of three representative conformations in **c** and their corresponding density maps in side view (left) and back view (right). The NBD dimer is colored in yellow and the RING ZnA domain in pink. All other chains are set to 100% transparency to emphasize regions of interest. **e,** Predicted all-atom structures of the RAG SEC complex by CryoDyna-CG and backmapping. The left and right panels correspond to conformations #1 and

#9 from the PC1 trajectory in Fig. S7a, respectively. During the lever-like motion of the RAG1 dimer, interchain hydrogen-bond interactions are formed (inset). Red, blue, and white indicate O, N, and H atoms, respectively.

## CryoDyna accurately partitions major states of translocating ribosome in a one-shot manner

Ribosomal translocation is a multi-subunit coordinated molecular event where EF-G hydrolyzes GTP to drive the movement of tRNAs along mRNA[36]. To model the complex mode of GTPase-powered ribosome-tRNA movement, we curated a dataset[27] of 58,433 particle images representing ribosome complexes after EF-G incorporation into the system (4681 nucleotides in 6 chains and 6656 residues in 53 chains;EMPIAR-10792). The crystal structure in H1 state (PDB:7PJV) is used as the initial structure to perform training process.

Uniform manifold approximation and projection (UMAP) analysis on the latent space of CryoDyna clearly separated these particles into three distinct clusters (Fig. 5a). These clusters correspond closely to the reported labels after multi-round 3D classifications[27]. We employed K-means clustering on the latent embeddings to sample 20 representative conformational states that collectively demonstrate the recovered multi-scale dynamics of functional domains. Structural analysis reveals that GTP hydrolysis during the transition from the H1/H2 to the CHI1 state drives large-scale conformational rearrangements of the LSU (50S subunit) and SSU (30S subunit), EF-G displacement, and tRNAs translocation (Fig. 5b-c, Supplementary Video 5). At the global scale, SSU head exhibits a counterclockwise rotation coupled with clockwise rotation of its body, which drives EF-G to rotate counterclockwisely toward the decoding center of the tRNA–mRNA complex. Concurrently, the LSU rotates clockwisely. At a finer scale, movements within the ribosomal RNA and EF-G promotes out-of-plane rotation of the tRNAs, relocating them within the SSU body from the A/P sites to the P/E sites. As the complex transitions from the CH1 to the CH2 state (Fig. 5b-c), CryoDyna further captured counterclockwise rotations of the LSU and SSU head, along with continued ingress of EF-G into the inter-subunit space formed by the SSU head and the tRNA–mRNA complex. Both tRNAs undergo a counterclockwise rotation in concert with the movements of the ribosomal subunits.

The additional density reconstruction using the learned latent variables revealed consistent multi-scale motions (Fig. 5d-e). Visualization of the P-site fMet-tRNA, A-site fMet-Phe-tRNA, and EF-G positions within the predicted density maps demonstrated motion patterns congruent with the structural model (Fig. 5e). Compared with the latent space learnt by 3DFlex[15],

CryoDyna additionally partitions CHI2 state out of the CHI states. Moreover, CryoDyna generated more consistent classification results with published labels than CryoSTAR (Adjusted Rand Index: 0.937 vs 0.727) after fitting a 3-component gaussian mixture model on the latent space (Fig. S8). These improvements led to a higher consistency between the experimental density map and the homogenous reconstructed map of the CHI2 state using particles assigned by CryoDyna (Fig. 5f). Above analysis demonstrated CryoDyna's ability and efficiency of automatically uncovering different functional states in a one-shot manner, in contrast to the traditional multi-round 3D classification construction protocol.

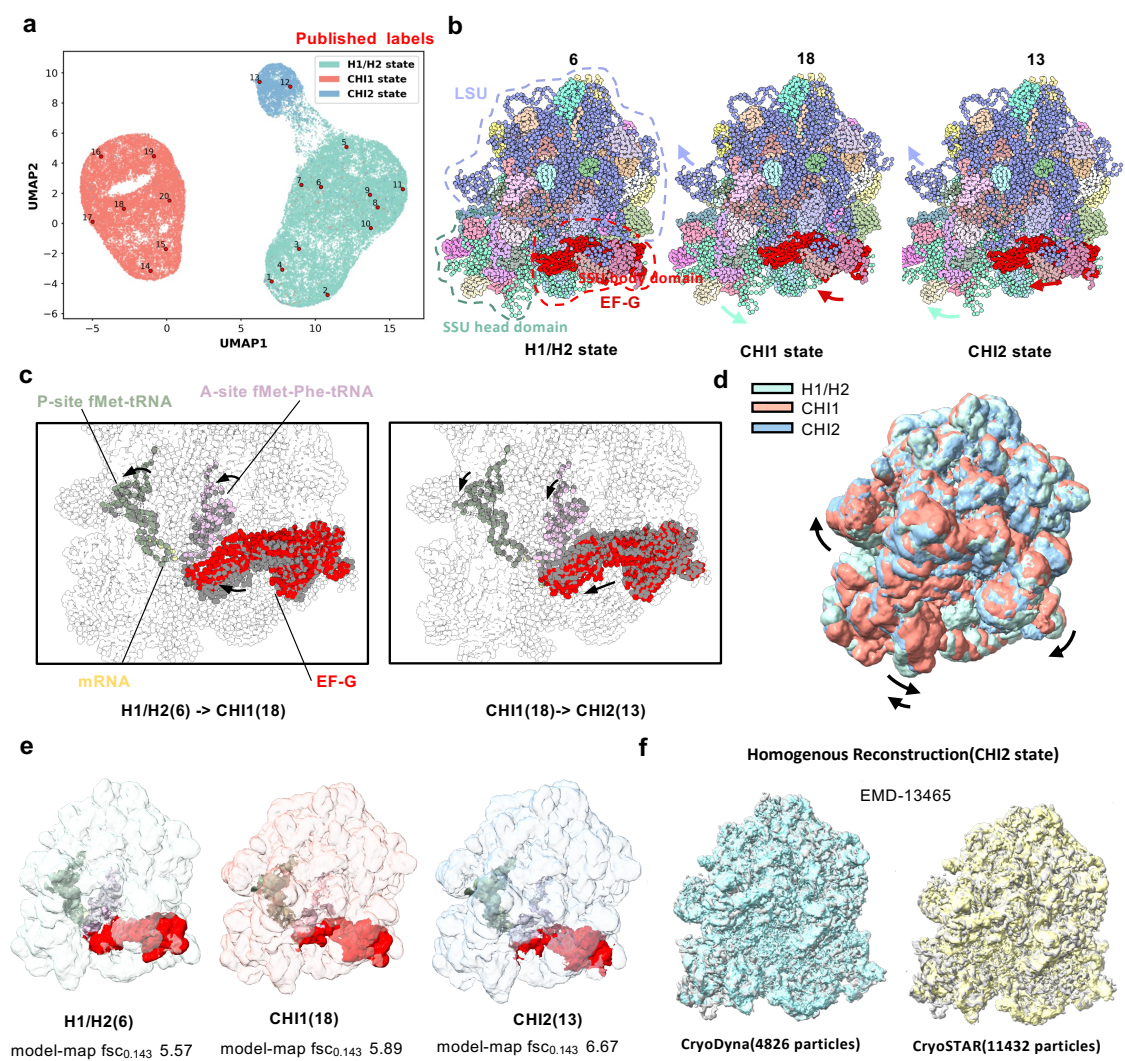

**Fig. 5 | CryoDyna accurately partitions different functional states of translocating ribosome (EMPIAR-10792). a**, UMAP visualization of the cryoDyna latent space, colored by major states assigned from the 3D classification in ref[27]. Twenty representative cluster centers by K-means Clustering are colored in red. **b**, Representative structures labeled in **a** for H1/H2 state (left), CHI1 state (middle), and CHI2 state (right), respectively. **c**, Enlarged visualization of the interaction area of tRNA-mRNA-EF-G. Left panel denotes the transition from H1/H2 state to CHI1 state where structure in H1/H2 state are colored in grey and P-site fMet-tRNA, A-site fMet-Phe-tRNA, mRNA and EF-G in CHI1 state are colored in celadon, pink, yellow and red, respectively. Right panel denotes the transition from CHI1 state to CHI2 state where structure

in CH1 state are colored in grey and P-site fMet-tRNA, A-site fMet-Phe-tRNA, mRNA and EF-G in CHI2 state are colored in celadon, pink, yellow and red, respectively. All other chains are set to 100% transparency to emphasize regions of interest. **d**, Superposition of density maps predicted by volume decoder corresponding to the same particle image in **b** in different states. **e**, Density maps predicted by volume decoder corresponding to the same particle image in **b** with corresponding density for P-site fMet-tRNA, A-site fMet-Phe-tRNA, mRNA and EF-G colored in celadon, pink, yellow and red, respectively. Model-map $FSC_{0.143}$ measures the consistency of atomic structure and corresponding density map. All density maps are shown using the same isosurface level. **f**, Superposition of the density map (blue) by homogeneous construction using 4826 particles assigned to CHI2 state by CryoDyna and deposited density map (grey) in CHI2 state. Map-map $FSC_{0.143}$ measures the consistency of experimental density map and constructed density map.

## CryoDyna models the stepwise closure of AP-1:Arf1:tetherin-Nef trimer

HIV-1 Nef protein evades immune defenses by hijacking the cell's transport system (AP-1/Arf1) to trap the antiviral protein tetherin in the Golgi and prevent it from blocking new virus particles release in the infected cells[37]. Using cryo-EM, Kyle et al.[28] resolved the AP-1:Arf1:tetherin-Nef closed trimer structure through subparticle extraction of monomeric subunits followed by flexible fitting into the full trimer density map, at the expense of eliminating dynamic signals for one of the subunits (Fig. 6a). We aim to recover here the hidden motion for the dynamic monomer from the original image stacks. We first fitted their resolved monomer subunit (PDB:6CM9) into the density map (EMDB-7458) to build the initial structure and reuse the poses determined by Kyle et al[28]. Notably, in the resolved structure, Nef contains only the core domain that interacts with the AP-1 μ subunit and tetherin, as well as the LL motif binding to the AP-1 σ1 subunit. In this study, we split them into two separate chains for ease of processing. The system contains 27 protein chains with 6582 residues in total.

We sampled 10 structures along PC1 in the latent space (Fig. 6b), with four representative structures shown in Fig. 6c. The PC1 trajectory reveals the transition of an AP-1:Arf1:tetherin-Nef monomer from an open to a closed trimer state: initially, the monomer interacts with the dimer solely via γArf1:57–61–γArf1:146–150 interaction and the Nef LL motif–AP-1 σ1 interface. Subsequently, its γArf1 approaches the third monomer's γArf1, while the Nef core domain shifts downward and to the right toward the third monomer's AP-1 σ1. The dynamic monomer's initial flattened conformation (with AP-1 γ-head/AP-1 σ1 distant from AP-1 β1/AP-1μ1/βArf1) evolves into a compact state through out-of-plane rotations of AP-1 β1/AP-1 μ1/βArf1 and in-plane rotations of AP-1 γ-head/AP-1 σ1, leading to a similar final conformation with the other two monomer subunits (Fig. S9a, Supplementary Video 6). A stepwise reconstruction of the open-closed state transition for the trimer is thus established. The predicted density maps by volume decoder showed recognized secondary structure features in partial region of the dynamic monomer (Fig. 6d). They are generally compatible with the predicted structure transition trajectory (Fig. S9b-c). Next, we validated these motions by homogenous reconstruction in CryoDRGN using particle subsets partitioned by performing a 5-component Gaussian mixture model fitting on the PC1 latent space (Fig. S9d). The reconstructed maps showed good agreement with the predicted molecule motions spanning open, intermediate, and closed states (Fig. S9e). The dynamic monomer exhibits a block-wise motion pattern, characterized by opposed movement between the AP-1 γ‑head/σ1 and AP-1 β1/μ1/βArf1 modules. This conformational shift generates transient inter-subunit cavities, which may represent novel therapeutic targets to inhibit closed trimer assembly and promote the delivery of antiviral protein tetherin to the cell membrane.

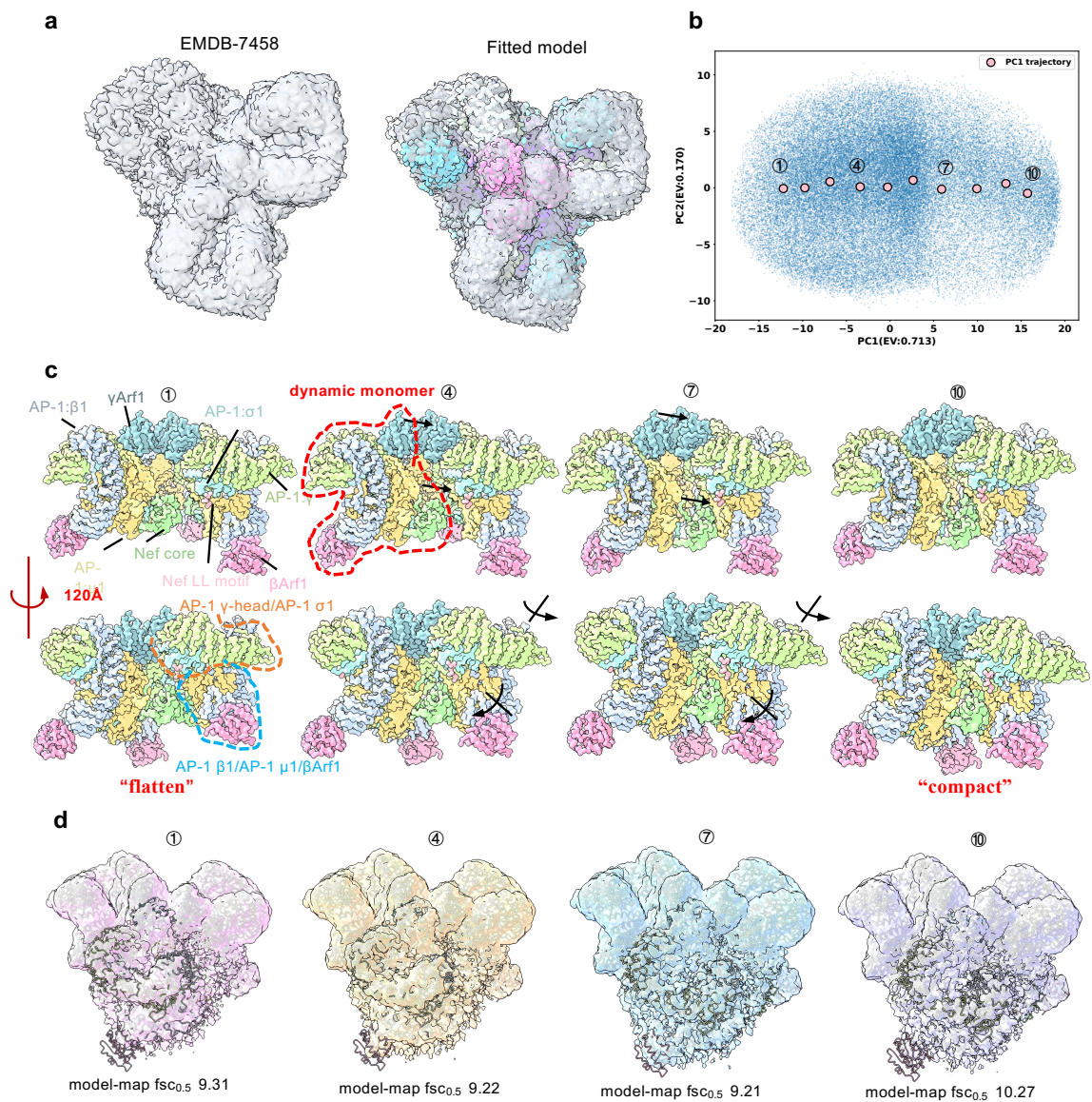

**Fig. 6 | CryoDyna models stepwise closure of AP-1:Arf1:tetherin-Nef trimer (EMPIAR-10177). a**, Visualization of deposited density map (EMDB-7458) for AP-1:Arf1:tetherin-Nef trimer (left) and Superposition of the fitted model with the deposited map (right). **b**, PCA visualization of the CryoDyna latent space. EV, explained variance. Ten representative conformations sampled along PC1 are colored in pink. **c**, Visualization of four representative structures labeled as 1,4,7,10 in **b** (left to right). The dynamic monomer is enclosed by the red dotted line. Two blocks which move independently are enclosed by orange and lightblue dotted lines, respectively. **d**, Superposition of four representative structures labeled as 1,4,7,10 in **b** with their predicted density maps by volume decoder. Model-map $FSC_{0.5}$ measures the consistency of atomic structure and corresponding density map. All density maps are shown using the same isosurface level.

# Discussion

CryoDyna offers an end-to-end solution for directly inferring macromolecular structural dynamics from 2D particle images. Compared with previous heterogeneity reconstruction methods that integrates residue-level models, our approach fuses multi-view image information to prevent hallucinated structure induced by signal from single image. In addition, we designed a hierarchical module to correlate residue deformation with its local environment, thereby takes into physical interactions in producing structural dynamics. On two simulated datasets, CryoDyna consistently generated structures align well with the native conformations. Notably, it also achieved robust and reliable performance in predicting conformations that deviate substantially from the initial structures. On four experimental datasets, CryoDyna was used to resolve continuous structural heterogeneity for yeast spliceosomal U4/U6.U5 tri-snRNP and AP-1:Arf1:tetherin-Nef trimer, yielding reliable trajectories for structural transition. For the RAG signal end complex, integration of structure prediction models like AF3 with CryoDyna effectively assists to analyze dynamics in regions unresolved by traditional reconstruction approaches. CryoDyna also enables the identification of major functional states of translocating ribosome in one-way forward training, which would otherwise require several rounds of human-supervised 3D classifications in traditional protocols.

In addition, we showed that the expanded bead-level reconstruction and all-atom back-mapping of CryoDyna can enable unprecedented resolution in characterizing conformational heterogeneity at fine scales. This strategy not only enhances modeling accuracy but also provides unique understanding of mechanisms requiring high-resolution structural features, allowing us to trace subtle yet functionally critical changes like the motion of phosphor-ribose backbone of nucleic acid chains (Fig. 3f) and analyze the interaction forces driving structural transformation (Fig. 4g). We hope that the backmapped all-atom structures can further provide opportunities to capture complex biological mechanisms and to integrate particle image data with molecular dynamics simulations, for example, by serving as intermediate states or extracting collective variable (CV) to perform enhancing sampling.

Although CryoDyna has been validated across multiple datasets, several limitations remain. First, for systems with particularly complicated structural heterogeneity, the initial pose estimation from upstream homogeneous reconstruction may lack sufficient accuracy, limiting CryoDyna interpretations for particles with large pose deviations. Joint optimization of poses and conformations by integrating CryoDyna with the two-stage pose determination pipeline implemented in CryoDRGN-AI[38] can be crucial for analyzing such systems. Secondly, CryoDyna does not aim to handle compositional heterogeneity due to its reliance on a single consensus reference structure. To address this issue, a chain-mask strategy can be introduced to explicitly mask partial chains from the whole system. Thirdly, extending the Gaussian-based representation to small molecules, lipids, and carbohydrates would facilitate observation of a broader spectrum of dynamic conformations.

# Methods

## The architecture of CryoDyna

### Encoder-decoder structure of CryoDyna

CryoDyna employs an encoder-decoder architecture to predict coordinate displacements ($\Delta S \in \mathbb{R}^{N \times 3}$) relative to the initial reference structure $S \in \mathbb{R}^{N \times 3}$. The encoder module processes an input 2D particle image $I \in \mathbb{R}^{D \times D}$, where $D$ refers to the dimension of the image, through a multilayer perceptron (MLP), generating a low-dimensional latent representation $Z$. For training stability, we use SiLU as activation function and add layer normalization between consecutive linear layers.

A cross-view attention mechanism merges image information from other images in a batch to update the latent code $Z$. To avoid over-smoothness of the latent code, we use TransformerConv operator supplemented in Pytorch Geometric[39] where representation of itself and aggregated information from other images are multiplied by different matrices defined as in Equation (1), respectively:

$$Z'_i = W_1 Z_i + \sum_{j \in \mathcal{N}(i)} \alpha_{i,j} W_2 Z_j \qquad (1)$$

$\alpha_{i,j}$ here represents attention coefficient. We construct a fully connected graph in a training batch and use one-layer cross-view attention in predicting $C_\alpha$ displacements.

Next, a hierarchical deformation decoder transforms the latent code **Z** to coordinate displacements $\Delta S$. In detail, a linear layer converts latent code **Z** to deformation representation $H \in \mathbb{R}^{N \times W}$, where **N** denotes the residue number and **W** denotes the dimension of deformation representation. We employ the 'mkdssp' command in DSSP[40] to assign the Secondary Structural Elements (SSEs) of the initial structure. Consecutive residues sharing identical secondary structure are merged into a single SSE block. Disordered regions lacking in consecutive identical SSEs are divided into SSE blocks of at least three residues. Residue-level representations from the same SSE block will be aggregated into a meta-node representation along with the positional vector, as in Equation (2):

$$H_{meta_j} = \sum_{i \in meta_j} MLP\left(concat\left(H_i, PV_{i \to meta_j}\right)\right) \qquad (2)$$

where $PV_{i \to meta_j}$ denotes the positional vector from residue node to meta-node derived from the initial structure. Graph attention is then performed on the meta-node representation within K-nearest neighbors, which can be written as:

$$H'_{meta_j} = \sum_{meta_k \in KNN(meta_j)} \alpha_{i,j} W_3 H_{meta_k} \qquad (3)$$

The updated meta-node representation is broadcasted to corresponding residue-level representation using a gating fusion mechanism:

$$H'_i = MLP\left(concat\left(H_i, H'_{meta_j}, PV_{meta_j \to i}\right)\right) * H'_{meta_j} + H_i \qquad (4)$$

where $PV_{meta_j \to i}$ denotes the positional vector from meta-node to residue node derived from the initial structure. To encourage cross-SSE motions, nearest four meta-nodes are also used to update residue-level representation, which can be written as:

$$H''_i = \sum_{meta_n \in KNN(i)} MLP(concat(H_i, H'_{meta_n}, PV_{meta_n \to i})) * H'_{meta_n} + H'_i \qquad (5)$$

We stack two layers of multi-scale deformation module and use the updated residue representation as readout to predict coordinate update through a three-layer MLP. It is noteworthy that while the image encoder employs SiLU as activation functions, all other components in the network utilize ELU to accelerate convergence. Once the training completes, the conformation corresponding to each image can be generated by passing its deformation code through the decoder.

**Hyperparameter setups in CryoDyna**

The image encoder follows a default architecture with progressively decreasing MLP dimensions (D² → 512 → 256 → 128), where cross-view attention operates on 128-dimensional latent variables. They are then compressed to 64 dimensions, before a linear layer transforms them into deformation codes of size num_pts × 32. All subsequent graph neural network updates are performed in this 32-dimensional space.

**Architecture modification in CryoDyna-CG**

Residue-level and Bead-level reconstruction adopts the same hyperparameters for network architecture. For bead-level prediction, we introduce a bead-level representation in the multi-scale deformation decoder. During the aggregation stage, features from beads belonging to the same residue are fused into a unified residue-level representation first as the same operator defined in Equation (3). In the broadcast stage, residue-level representations are used to update the representation of its constituent beads through the same gating mechanism as defined in Equation (4), and the updated bead features are used to predict per-bead deformations.

**Gaussian electron density simulation**

To establish correspondence between deformed structures and particle images, we first transform the deformed structure $\tilde{S}$ to a set of Gaussian functions simulating a three-dimensional electron density volume $V \in \mathbb{R}^{D \times D \times D}$. These volumes are then projected using predetermined orientations to generate corresponding simulated particle images. CryoDyna models the electron density of each residue using a Gaussian function and computes the volumetric density through summation. This process can be written as Equation (6):

$$\hat{V}(x) = \sum_i^N A_i \exp\left(-\frac{\|X-\mu_i\|^2}{2\sigma_i^2}\right) \tag{6}$$

where $X \in \mathbb{R}^3$ denotes sampled spatial grid coordinate, $\mu_i$ is the predicted $C_\alpha$ coordinate of the $i$-th residue, $\sigma_i$ refers to the radius of the $i$-th residue (set to 2 Å empirically), and the amplitude $A_i$ counts the total electrons of the $i$-th residue.

For bead-level prediction, $\mu_i$, $\sigma_i$ and $A_i$ represent the coordinate, effective radius (following Lennard-Jones (LJ) parameters in MARTINI), and total number of electrons of the $i$-th bead in the reference structure. Compared to residue-level prediction which uses a uniform radius for all residues, assigning different radii based on bead types in CryoDyna-CG can yield a more accurate representation of electron distribution.

**Image formation model**

For a volumetric density $\hat{V}$ simulated from the deformed structure, we generate the corresponding 2D projections as following:

$$I_{pred} = (CTF \circ \Pi_{2D}^t \circ R)(\hat{V}) \tag{7}$$

where $R \in SO(3)$ is the rotation matrix, $\Pi_{2D}^t$ is a 2D in-plane translation operator with translation vector $t \in \mathbb{R}^2$, and CTF is the contrast transfer function. All poses used in this study are predetermined by upstream reconstruction pipeline.

**Structure regularization**

**Residue-level regularization**

For residue-level predictions, we use continuity loss, clash loss and Elastic Network (EN) loss in the same form as CryoSTAR[18] to preserve structural validity, which can be written as:

$$\mathcal{L}_{continuity} = \frac{1}{N-1}\sum_{i=1}^{N-1}|d_{i,i+1}^0 - d_{i,i+1}|^2 \tag{8}$$

$$\mathcal{L}_{clash} = \frac{1}{|P_{clash}|}\sum_{(i,j)\in P_{clash}}|d_{i,i+1} - k_{clash}|^2 \tag{9}$$

$$\mathcal{L}_{EN} = \frac{1}{|P'_{EN}|}\sum_{(i,j)\in P'_{EN}}|d_{ij}^0 - d_{ij}|^2 \tag{10}$$

$$P'_{EN} = \{(i,j) \mid (i,j) \in P_{EN},\ Var(d_{ij}) < \text{the}\ p\text{-th percentile of } \mathcal{V}\} \tag{11}$$

$\mathcal{L}_{continuity}$ prevents chain break in the optimization process where $d_{i,i+1}^0$ and $d_{i,i+1}$ refers to the distance of adjacent residue pair in the initial structure and the predicted structures. $\mathcal{L}_{clash}$ penalizes colliding residue pairs in the predicted structure where $P_{clash}$ represents nonadjacent residue pairs whose distances fall in 4-10 Å in initial structure but are closer than 4 Å in the predicted structure. $k_{clash}$ is set to 4 Å. $\mathcal{L}_{EN}$ keeps local rigidity where $P_{EN}$ denotes non-adjacent residue pairs within 12 Å while $P'_{EN}$ is a dynamic subset of $P_{EN}$, retaining only those edges whose distance change falls below the a specific threshold in each training step. This regulation of residue model effectively relaxes edges that exhibit potential flexibility based on image signal. It's noted that we only add EN loss on all intra-chain edges (threshold:0.99, 0.8 for adenylate kinase) and inter-chain edges (threshold:1.0) between nucleic acids.

**Bonded regularization for CryoDyna-CG**

For bead-level predictions, we design a composite regularization strategy that integrates MARTINI version 2.2 force field-derived bonded potentials including bond lengths and angles, dihedral constraints, and non-bonded Morse potential with EN loss derived from the initial structure.

**Bond** The MARTINI force field includes two main types of bonds: backbone bond and sidechain bond. Backbone bonds connect two adjacent backbone beads, while sidechain bonds link either a backbone bead and a sidechain bead within the same subunit (amino acid or nucleotide), or two sidechain beads. Backbone bonds for proteins are further categorized based on secondary structure into: constraints, short elastic bonds for extended regions, long elastic bonds for extended regions, and others. For MARTINI CG structures, a bond loss similar to the harmonic bond potential is used to constrain bead connectivity, defined as:

$$\mathcal{L}_{bond} = \frac{1}{|P_{bond}|} \sum_{(i,j) \in P_{bond}} \frac{k_{b(i,j)}}{k_0} |d_{b(i,j)} - d_{i,j}|^2 \qquad (12)$$

Here, $P_{bond}$ denotes the set of bead pairs forming bonded interactions in the MARTINI force field. $d_{i,j}$ is the distance between beads $i$ and $j$ in the predicted structure, while $k_{b(i,j)}$ and $d_{b(i,j)}$ represent the force constant and equilibrium bond length for the pair, respectively. To scale the potential, all force constants are normalized by a base value $k_0$ ($1250\ kJ \cdot mol^{-1}$). Bead pairs modeled with harmonic potential energy use their corresponding force field parameters in the loss computation, while those treated as constraints are assigned a relatively large force constant of $100{,}000\ kJ \cdot mol^{-1} \cdot nm^{-2}$ .

**Angle** The MARTINI force field includes three types of protein angles: the angle between three consecutive backbone beads, the angle formed by two neighboring backbone beads and the first sidechain bead of an amino acid, and the angle formed by a backbone bead and two consecutive sidechain beads of the same amino acid. The angle loss is defined as:

$$\mathcal{L}_{angle} = \frac{2}{N_{bead}^2} \sum_{(i,j,k) \in P_{angle}} \frac{k_{a(i,j,k)}}{k_0} |cos\theta_{a(i,j,k)} - cos\theta_{i,j,k}|^2 \tag{13}$$

Here, $P_{angle}$ denotes the set of bead triplets forming angles in the MARTINI force field. $\theta_{i,j,k}$ is the angle formed by beads $i$, $j$, and $k$ in the predicted structure, while $k_{a(i,j,k)}$ and $\theta_{a(i,j,k)}$ represent the force constant and equilibrium angle for the triplet, respectively. $N_{bead}$ denotes the total number of beads in the system.

**Dihedral** In the MARTINI force field, dihedrals are classified into two types: proper dihedrals, which maintain the secondary structure of the peptide backbone, and improper dihedrals, which prevent out-of-plane distortions of planar groups. The corresponding losses are defined as:

$$\mathcal{L}_{dihedral} = \frac{2}{N_{bead}^2} \sum_{(i,j,k,l) \in P_{dihedral}} \frac{k_{d(i,j,k,l)}}{k_0} [1 + \cos(n\phi_{i,j,k,l} - \phi_{d(i,j,k,l)})] \tag{14}$$

$$\mathcal{L}_{improper} = \frac{2}{N_{bead}^2} \sum_{(i,j,k,l) \in P_{improper}} \frac{k_{id(i,j,k,l)}}{k_0} |\phi_{id(i,j,k,l)} - \phi_{i,j,k,l}|^2 \tag{15}$$

Here, $P_{dihedral}$ and $P_{improper}$ denote the sets of bead quadruplets forming proper and improper dihedrals in the MARTINI force field, respectively. $\phi_{i,j,k,l}$ is the dihedral angle formed by beads $i$, $j$, $k$, and $l$ in the predicted structure. $k_{d(i,j,k,l)}$ and $k_{id(i,j,k,l)}$ represent the force constants for proper and improper dihedrals, respectively, while $\phi_{d(i,j,k,l)}$ and $\phi_{id(i,j,k,l)}$ are the equilibrium dihedrals for beads $i$, $j$, $k$ and $l$. $n$ in Equation (14) denotes the multiplicity. For nucleic acids, each nucleotide is mapped to six or seven CG beads, and bonded potentials are applied to the system in a similar way.

**Nonbonded regularization for CryoDyna-CG**

**Nonbonded Morse potential** To prevent clashes between beads and keep nonbonded interactions, a loss term in the form of a Morse potential is applied:

$$\mathcal{L}_{Morse} = \frac{2}{P_{Morse}} \sum_{(i,j) \in P_{Morse}} \frac{\epsilon_{i,j}}{k_0} \| 1 - e^{-a(d_{i,j} - d_{m(i,j)})} \|^2 \tag{16}$$

$$d_{m(i,j)} = 2^{\frac{1}{6}}\sigma_{i,j} \tag{17}$$

Here, $P_{Morse}$ denotes all bead pairs without bonded constraints, considering only those within 12 Å in the initial structure. $\epsilon_{i,j}$ and $\sigma_{i,j}$ are set according to LJ potential parameters from the MARTINI force field, and $d_{m(i,j)}$ is set to ensure that the Morse potential shares the same energy minimum as the LJ potential. The parameter $a$ is set to 1. Unlike the clash loss which pulls close-contact pairs toward a fixed distance, this loss accounts for the interactions between different bead types by referencing the LJ potential.

**Elastic network** To further maintain the local structural rigidity, EN loss terms consistent with the residue-level predictions of CryoDyna are applied. In CryoDyna-CG, all intra-chain edges, inter-chain edges between nucleic acids and proteins, and inter-chain edges between nucleic acids are applied with a distance threshold of 0.99 (0.8 for adenylate kinase), 0.99 and 1.00, respectively.

## Loss function

For residue-level prediction, the loss function consists of two components: (1) image reconstruction loss $\mathcal{L}_{image}$, (2) structural regularization loss $\mathcal{L}_{reg}$.

$$\mathcal{L}_{total} = \mathcal{L}_{image} + \mathcal{L}_{reg} \tag{18}$$

CryoDyna employs a correlation-based loss to align 2D projections to raw images, which can be written as:

$$\mathcal{L}_{image} = -\frac{I_{pred} * I_{raw}}{\|I_{pred}\| * \|I_{raw}\|} \tag{19}$$

The structural regularization loss comprises three components defined in Equation (8)-(11):

$$\mathcal{L}_{reg} = \mathcal{L}_{continiuty} + \mathcal{L}_{clash} + \mathcal{L}_{EN} \tag{20}$$

For bead-level CryoDyna-CG prediction, structural regularization $L_{reg}^{cg}$ contains bonded and nonbonded terms defined in Equation (12)-(17):

$$L_{reg}^{cg} = \mathcal{L}_{bond} + \mathcal{L}_{angle} + \mathcal{L}_{dihedral} + \mathcal{L}_{improper} + \mathcal{L}_{Morse} + \mathcal{L}_{EN} \tag{21}$$

Similar to the loss design at the residue level, $\mathcal{L}_{bond}$ maintains the connectivity between adjacent beads, $\mathcal{L}_{Morse}$ prevents bead–bead clashes, and $\mathcal{L}_{EN}$ adopts the same form.

Meanwhile, $\mathcal{L}_{angle}$, $\mathcal{L}_{dihedral}$, and $\mathcal{L}_{improper}$ ensure the geometric consistency of local structures. Therefore, final loss function for bead-level prediction can be written as:

$$\mathcal{L}_{total}^{cg} = \mathcal{L}_{image} + L_{reg}^{cg} \tag{22}$$

## Coarse-grained model construction

The MARTINI 2.2 force field[21–23] is utilized to generate CG representations of the biomolecular systems including nucleic acids and proteins. This representation maps approximately four heavy atoms to a single CG bead (4:1 mapping ratio), while ring-like fragments are represented at a higher resolution. For DNA and RNA, each nucleotide is mapped to 6–7 beads, with the backbone represented by three beads corresponding to the phosphate group (1 bead) and the sugar moiety (2 beads). To describe interactions, beads are categorized into four main types: polar (P), nonpolar (N), apolar (C), and charged (Q), each further subdivided based on hydrogen-bonding capability or degree of polarity. The strength of non-bonded interactions depends on bead types, reflected in differences in their effective sizes and Lennard-Jones well depths. Before training of CryoDyna-CG, reference structure is first mapped to CG coordinates and energy-minimized by the steepest descent algorithm using GROMACS[41] (version 2023) . It is then provided to the model as structural priors.

## Training and optimization

All of the experiments in this study were performed on a single NVIDIA Tesla A100 graphics processing units (GPUs). The neural network parameters were initialized using PyTorch's default random initialization scheme. We employed the Adam optimizer for gradient descent with a fixed learning rate of $1\times10^{-4}$ throughout training. The same setup were used for CryoDyna and CryoDyna-CG.

## Recover all-atom structure from coarse-grained model

To reconstruct all-atom structures from CryoDyna-generated CG models, we employ a hybrid reconstruction strategy combining fragment-based and geometric-based approaches. For protein and DNA components, atomic structures are backmapped using the CG2AT2[30] method, where fragments are aligned to CG beads. RNA components are converted to atomistic structures through the geometric-based Backward[31] protocol. These reconstructed structures go through chain-wise energy minimization followed by a full-system relaxation and a 5-ps NVT simulation with positional restraints. Molecular dynamics simulations are performed using GROMACS[41] with the CHARMM36[42] force field.

## Validation

### Volume decoder

All analyses on experimental datasets in this study were validated by a volume decoder. When the training of CryoDyna completes, the 64-dimensional latent variables are retained to learn a mapping from each latent variable to a density map that corresponds to its original input image. We utilize a NeRF-like architecture[43] where the latent variable is concatenated with a 3D grid coordinate to predict the density value of this grid through a five-layer MLP. The hidden dimension for each layer is set to 1024. The input coordinates are encoded with a positional embedding of 3*D sinusoids (D denotes the side length of the image), each with a Gaussian-distributed frequency. During training, for each particle, we extract the central slice of the density volume determined by particle pose. A mean squared error (MSE) loss is optimized between the predicted densities on corresponding grid points for the central slice and the particle images. For volume reconstruction during inference, the trained model predicts densities of all 3D grid points of the volume to assemble the complete 3D map.

### Cluster reconstruction in CryoDRGN

For the translocating ribosome and the AP-1:Arf1:tetherin-Nef trimer, we performed cluster reconstruction to further validate the separated states. We picked subsets of particles corresponding to a specific state and used the "train_nn.py" scripts in CryoDRGN[12] (version 3.4.4) for homogenous reconstruction to validate each state. The training were performed for 50 epochs.

### CryoSTAR

We installed CryoSTAR locally. All experiments involving CryoSTAR in this study used the default configuration, with hyperparameters (learning rate, batch size, downsampled image size) consistent with those of CryoDyna, except that on the SARS-CoV spike protein dataset, we increased the batch size to 128 to enhance training stability.

## Two-stage atomic structure reconstruction using CryoDRGN and Phenix

For the simulated spike dataset, we performed a two-stage atomic structure reconstruction with CryoDRGN and Phenix[33] to benchmark the end-to-end reconstruction quality. We first used heterogenous reconstruction function implemented in CryoDRGN (version 3.4.4) to analyze the simulated spike dataset with the default setting and sampled 100 density maps corresponding to picked images. Then, the initial structure was refined according to the sampled maps in phenix-1.21.2-5419. In detail, we docked the initial structure into the maps using the "phenix.dock_in_map" command which was followed by 50 cycles of "phenix.real_space_refine" . Above operations used the default parameters.

## Dataset

The basic information, training parameters, and running time for each dataset used in this study are comprehensively summarized in Supplementary Table 1.

**Simulated datasets**

**Adenylate kinase dataset** This synthetic dataset was collected from Ref[18] which performed a structural interpolation between open (PDB:4AKE) and closed state (PDB:1AKE) atomic models, comprising 50,000 (1000 images for each interpolated structure) simulated particle images with an image size of 128 ×128 and a pixel size of 1 Å/pix. The signal-to-noise ratio (SNR) is 0.001.

**SARS-Cov-2 Spike dataset** This dataset was collected from Jeon et al[24] who analyzed a long-timescale molecular dynamics trajectory of activating SARS-Cov-2 spike protein. The dataset finally contains 100,000 particle images at a SNR of 0.1 with an image size 256 × 256 and a pixel size of 1.5 Å/pix.

**Experimental datasets**

**Yeast spliceosomal U4/U6.U5 tri-snRNP** The EMPIAR-10073 dataset[25] was downsampled to 128 × 128 from the original image size of 380 × 380, giving a pixel size of 4.15625 Å/pix. We process the full dataset of deposited particles (138,899 particles).

**RAG signal-end complex** The EMPIAR-10049 dataset[26] used the original image size of 192 × 192 with a pixel size of 1.23 Å/pix. We processed the full dataset of deposited particles (108,544 particles). We used AlphaFold3 server (https://alphafoldserver.com/) in default setting to generate the full system of RAG signal-end complex except the disordered region of RAG2 PHD domain. The predicted sequence was derived from PDB:3JBX (nicked 12-RSS/23-RSS;RAG1/2), PDB:3JBW (partial 12-RSS/23-RSS) and PDB:2GZK (HMGB1 box 10-48).

**Translocating ribosome** The EMPIAR-10792 dataset[27] was downsampled to 144 × 144 from the original image size of 288 × 288, giving a pixel size of 2.32 Å/pix. We merged all of the starfiles with EF-G present and led to a dataset with 53,821 particles.

**AP-1:Arf1:tetherin-Nef Trimer** The EMPIAR-10177 dataset[28] was downsampled to 256 × 256 from the original image size of 384 × 384, giving a pixel size of 1.6 Å/pix. We processed the full dataset of deposited particles (61,929 particles). The PDB:6CM9 atomic structure was docked in different region of EMDB-7458 in ChimeraX-1.6[44] and the final trimer structure acted as the initial structure .

## Evaluation Metrics

**Root-mean-square deviation (RMSD)** measures atomic-level structural differences after optimal superposition. In our analysis, we compute Cα-RMSD, bead-RMSD and all-atom RMSD for residue-level predictions, bead-level predictions, and structures after backmapping protocol, respectively.

**Model-map FSC** evaluates the agreement between an atomic model and its corresponding cryo-EM density map. We used the "e2proc3d.py --calcfsc" command implemented in EMAN2[45] to calculate the Fourier Shell Correlation (FSC) between density simulated from the atomic structure and the reconstructed particle density map. The reported model-map $FSC_{0.5}$ resolution corresponds to the spatial frequency where the FSC curve crosses the 0.5 threshold. Specifically, for the EMPIAR-10792 dataset, we report the $FSC_{0.143}$ due to approximation errors for abundant nucleic acids using single gaussian function. It's noted that the smallest resolvable detail in an image is fundamentally limited by twice the pixel size according to the Nyquist limit.

**Base-pair parameters** characterize the relative spatial relationship within a nucleotide pair. For the ten all-atom structures backmapped from CryoDyna-CG predictions in the 10049 dataset, we used CURVES+[46] to evaluate the 23-RSS DNA in terms of three translational parameters (Shear, Stretch, Stagger) and three rotational parameters (Buckle, Propeller, Opening).

**Base-step parameters** describe the relative position and orientation between adjacent base pairs. For the ten all-atom structures backmapped from CryoDyna-CG predictions in the 10049 dataset, we used CURVES+[46] to calculate one translational parameter (Rise) and one rotational parameter (Twist) for the 23-RSS DNA.

## Data availability

Input files, trained CryoDyna models, and generated structures will be released at https://osf.io/7vu5x/. Two simulated datasets can be obtained at https://drive.google.com/file/d/1OMvwBFUjR-97-nRMAcqXBfpNqEn5eFvw/view (Adenylate kinase dataset)[18] and https://zenodo.org/records/14941494 (SARS-Cov-2 Spike dataset)[24]. We used the following public experimental datasets: EMPIAR-10073 (cryo-EM data of the Yeast spliceosomal U4/U6.U5 tri-snRNP), EMPIAR-10049 (cryo-EM data of the RAG1–RAG2 signal end complex), EMPIAR-10792 (cryo-EM data of the translocating ribosome) and EMPIAR-10180 (cryo-EM data of the AP-1:Arf1:tetherin-Nef Trimer).

## Code availability

CryoDyna software is freely available at https://github.com/Qmi3/CryoDyna under the Apache License version 2.0.

## Acknowledgements

We thank Ma Liangwen from Peking University for her valuable suggestions regarding the processing of the electron microscopy datasets in this study. This work was supported by National Science and Technology Major Project (No. 2022ZD0115001), National Natural Science Foundation of China (92353304, T2495221) and New Cornerstone Science Foundation (NCI202305). This work was also supported by Changping Laboratory.

## Author contributions

Y.Q.G., and S.Liu conceived and supervised the project. C.Z., and S.Li developed the CryoDyna and CryoDyna-CG methods. C.Z., S.Li, Y.N., Z.Z. and S.Y. performed data collection, evaluation, and analysis. C.Z., S.Li, and Y.N. wrote the initial draft of the manuscript. All authors contributed ideas to the work and assisted in manuscript editing and revision.

## Competing interests

The authors declare no competing interests.

## References

1. Nogales, E. The development of cryo-EM into a mainstream structural biology technique. *Nat. Methods* **13**, 24–27 (2016).

2. Cheng, Y. Single-particle cryo-EM—How did it get here and where will it go. *Science* **361**, 876–880 (2018).

3. Brubaker, M. A., Punjani, A. & Fleet, D. J. Building Proteins in a Day: Efficient 3D Molecular Reconstruction. in *2015 IEEE Conference on Computer Vision and Pattern Recognition (CVPR)* 3099–3108 (2015). doi:10.1109/CVPR.2015.7298929.

4. Bonomi, M., Pellarin, R. & Vendruscolo, M. Simultaneous Determination of Protein Structure and Dynamics Using Cryo-Electron Microscopy. *Biophys. J.* **114**, 1604–1613 (2018).

5. Punjani, A., Rubinstein, J. L., Fleet, D. J. & Brubaker, M. A. cryoSPARC: algorithms for rapid unsupervised cryo-EM structure determination. *Nat. Methods* **14**, 290–296 (2017).

6. Scheres, S. H. W. *et al.* Disentangling conformational states of macromolecules in 3D-EM through likelihood optimization. *Nat. Methods* **4**, 27–29 (2007).

7. Scheres, S. H. W. Processing of Structurally Heterogeneous Cryo-EM Data in RELION. in *Methods in Enzymology* vol. 579 125–157 (Elsevier, 2016).

8. Scheres, S. H. W. RELION: Implementation of a Bayesian approach to cryo-EM structure determination. *J. Struct. Biol.* **180**, 519–530 (2012).

9. Tagare, H. D., Kucukelbir, A., Sigworth, F. J., Wang, H. & Rao, M. Directly reconstructing principal components of heterogeneous particles from cryo-EM images. *J. Struct. Biol.* **191**, 245–262 (2015).

10. Penczek, P. A., Kimmel, M. & Spahn, C. M. T. Identifying Conformational States of Macromolecules by Eigen-Analysis of Resampled Cryo-EM Images. *Structure* **19**, 1582–1590 (2011).

11. Heel, M. V., Portugal, R. V. & Schatz, M. Multivariate Statistical Analysis of Large Datasets: Single Particle Electron Microscopy. *Open J. Stat.* **06**, 701–739 (2016).

12. Zhong, E. D., Bepler, T., Berger, B. & Davis, J. H. CryoDRGN: reconstruction of heterogeneous cryo-EM structures using neural networks. *Nat. Methods* **18**, 176–185 (2021).

13. Zhong, E. D., Lerer, A., Davis, J. H. & Berger, B. CryoDRGN2: Ab initio neural reconstruction of 3D protein structures from real cryo-EM images. in *2021 IEEE/CVF International Conference on Computer Vision (ICCV)* 4046–4055 (IEEE, Montreal, QC, Canada, 2021). doi:10.1109/ICCV48922.2021.00403.

14. Chen, M. & Ludtke, S. J. Deep learning-based mixed-dimensional Gaussian mixture model for characterizing variability in cryo-EM. *Nat. Methods* **18**, 930–936 (2021).

15. Punjani, A. & Fleet, D. J. 3DFlex: determining structure and motion of flexible proteins from cryo-EM. *Nat. Methods* **20**, 860–870 (2023).

16. Hamitouche, I. & Jonic, S. DeepHEMNMA: ResNet-based hybrid analysis of continuous conformational heterogeneity in cryo-EM single particle images. *Front. Mol. Biosci.* **9**, 965645 (2022).

17. Herreros, D. *et al.* Estimating conformational landscapes from Cryo-EM particles by 3D Zernike polynomials. *Nat. Commun.* **14**, 154 (2023).

18. Li, Y., Zhou, Y., Yuan, J., Ye, F. & Gu, Q. CryoSTAR: leveraging structural priors and constraints for cryo-EM heterogeneous reconstruction. *Nat. Methods* **21**, 2318–2326 (2024).

19. Ducrocq, G., Grunewald, L., Westenhoff, S. & Lindsten, F. cryoSPHERE: Single-particle heterogeneous reconstruction from cryo EM. Preprint at https://doi.org/10.1101/2024.06.19.599686 (2024).

20. Monticelli, L. *et al.* The MARTINI Coarse-Grained Force Field: Extension to Proteins. *J. Chem. Theory Comput.* **4**, 819–834 (2008).

21. de Jong, D. H. *et al.* Improved parameters for the martini coarse-grained protein force field. *J. Chem. Theory Comput.* **9**, 687–697 (2013).

22. Uusitalo, J. J., Ingólfsson, H. I., Akhshi, P., Tieleman, D. P. & Marrink, S. J. Martini Coarse-Grained Force Field: Extension to DNA. *J. Chem. Theory Comput.* **11**, 3932–3945 (2015).

23. Uusitalo, J. J., Ingólfsson, H. I., Marrink, S. J. & Faustino, I. Martini Coarse-Grained Force Field: Extension to RNA. *Biophys. J.* **113**, 246–256 (2017).

24. Jeon, M. *et al.* CryoBench: Diverse and challenging datasets for the heterogeneity problem in cryo-EM. in *Advances in Neural Information Processing Systems* (eds Globerson, A. et al.) vol. 37 89468–89512 (Curran Associates, Inc., 2024).

25. Nguyen, T. H. D. *et al.* Cryo-EM structure of the yeast U4/U6.U5 tri-snRNP at 3.7 Å resolution. *Nature* **530**, 298–302 (2016).

26. Ru, H. *et al.* Molecular Mechanism of V(D)J Recombination from Synaptic RAG1-RAG2 Complex Structures. *Cell* **163**, 1138–1152 (2015).

27. Petrychenko, V. *et al.* Structural mechanism of GTPase-powered ribosome-tRNA movement. *Nat. Commun.* **12**, 5933 (2021).

28. Morris, K. L. *et al.* HIV-1 Nefs Are Cargo-Sensitive AP-1 Trimerization Switches in Tetherin Downregulation. *Cell* **174**, 659-671.e14 (2018).

29. Abramson, J. *et al.* Accurate structure prediction of biomolecular interactions with AlphaFold 3. *Nature* **630**, 493–500 (2024).

30. Vickery, O. N. & Stansfeld, P. J. CG2AT2: an Enhanced Fragment-Based Approach for Serial Multi-scale Molecular Dynamics Simulations. *J. Chem. Theory Comput.* **17**, 6472–6482 (2021).

31. Wassenaar, T. A., Pluhackova, K., Böckmann, R. A., Marrink, S. J. & Tieleman, D. P. Going Backward: A Flexible Geometric Approach to Reverse Transformation from Coarse Grained to Atomistic Models. *J. Chem. Theory Comput.* **10**, 676–690 (2014).

32. Wieczór, M., Tang, P. K., Orozco, M. & Cossio, P. Omicron mutations increase interdomain interactions and reduce epitope exposure in the SARS-CoV-2 spike. *iScience* **26**, 105981 (2023).

33. Afonine, P. V. *et al.* Real-space refinement in *PHENIX* for cryo-EM and crystallography. *Acta Crystallogr. Sect. Struct. Biol.* **74**, 531–544 (2018).

34. Liu, S., Rauhut, R., Vornlocher, H.-P. & Lührmann, R. The network of protein–protein interactions within the human U4/U6.U5 tri-snRNP. *RNA* **12**, 1418–1430 (2006).

35. Sadofsky, M. J. The RAG proteins in V(D)J recombination: more than just a nuclease. *Nucleic Acids Res.* **29**, 1399–1409 (2001).

36. Frank, J., Gao, H., Sengupta, J., Gao, N. & Taylor, D. J. The process of mRNA–tRNA translocation.

37. Coleman, S. H., Hitchin, D., Noviello, C. M. & Guatelli, J. C. HIV-1 Nef stabilizes AP-1 on membranes without inducing ARF1-independent de novo attachment. *Virology* **345**, 148–155 (2006).

38. Levy, A. *et al.* CryoDRGN-AI: neural ab initio reconstruction of challenging cryo-EM and cryo-ET datasets. *Nat. Methods* **22**, 1486–1494 (2025).

39. Fey, M. *et al.* PyG 2.0: Scalable Learning on Real World Graphs. Preprint at https://doi.org/10.48550/arXiv.2507.16991 (2025).

40. Hekkelman, M. L., Salmoral, D. Á., Perrakis, A. & Joosten, R. P. DSSP 4: FAIR annotation of protein secondary structure. *Protein Sci.* **34**, e70208 (2025).

41. Páll, S. *et al.* Heterogeneous parallelization and acceleration of molecular dynamics simulations in GROMACS. *J. Chem. Phys.* **153**, 134110 (2020).

42. Huang, J. *et al.* CHARMM36m: an improved force field for folded and intrinsically disordered proteins. *Nat. Methods* **14**, 71–73 (2017).

43. Mildenhall, B. *et al.* NeRF: Representing Scenes as Neural Radiance Fields for View Synthesis. Preprint at https://doi.org/10.48550/arXiv.2003.08934 (2020).

44. Goddard, T. D. *et al.* UCSF ChimeraX: Meeting modern challenges in visualization and analysis. *Protein Sci.* **27**, 14–25 (2018).

45. Tang, G. *et al.* EMAN2: An extensible image processing suite for electron microscopy. *J. Struct. Biol.* **157**, 38–46 (2007).

46. Blanchet, C., Pasi, M., Zakrzewska, K. & Lavery, R. CURVES+ web server for analyzing and visualizing the helical, backbone and groove parameters of nucleic acid structures. *Nucleic Acids Res.* **39**, W68–W73 (2011).